\documentclass[twocolumn,traditabstract,longauth]{aa}
\usepackage{amssymb}
\usepackage{mathptmx}
\usepackage{natbib}
\bibpunct{(}{)}{;}{a}{}{,}
\usepackage[]{graphicx}
\usepackage{lscape}
\usepackage{txfonts}
\usepackage{url}
\defcitealias{canameras15}{C15}
\defcitealias{canameras17}{C17}
\begin{document}

\title{\textit{Planck}'s dusty GEMS. IV. Star formation and feedback in a maximum starburst at \textit{z}=3 seen at 60-pc resolution
  \thanks{Based on ALMA data obtained with program 2015.1.01518.S}
  \thanks{Corresponding author: R.~Ca\~nameras, e-mail: canameras@dark-cosmology.dk}}
\author{R.~Ca\~nameras\inst{1}, N.~Nesvadba\inst{1}, R.~Kneissl\inst{2,3}, B.~Frye\inst{4}, R.~Gavazzi\inst{5}, S.~Koenig\inst{6}, E.~Le~Floc'h\inst{7}, M.~Limousin\inst{8}, I.~Oteo\inst{9,10}, D.~Scott\inst{11}}
\institute{
Institut d'Astrophysique Spatiale, CNRS, Univ. Paris-Sud, Universite Paris-Saclay, Bat.~121, 91405 Orsay Cedex, France
\and
European Southern Observatory, ESO Vitacura, Alonso de Cordova 3107, Vitacura,
Casilla 19001, Santiago, Chile
\and
Atacama Large Millimeter/submillimeter Array, ALMA Santiago Central Offices,
Alonso de Cordova 3107, Vitacura, Casilla 763-0355, Santiago, Chile
\and
Steward Observatory, University of Arizona, Tucson, AZ 85721, USA
\and
Institut d'Astrophysique de Paris, 75014, Paris, UPMC Univ. Paris 6, UMR7095
\and 
Chalmers University of Technology, Onsala Space Observatory, Onsala, Sweden 
\and
Laboratoire AIM, CEA/DSM/IRFU, CNRS, Universite Paris-Diderot, Bat.~709, 91191 Gif-sur-Yvette, France
\and
Aix Marseille Univ, CNRS, LAM, Laboratoire d'Astrophysique de Marseille, Marseille, France 
\and
Institute for Astronomy, University of Edinburgh, Royal Observatory,
Blackford Hill, Edinburgh, EH9 3HJ, UK
\and
European Southern Observatory, Karl-Schwarzschild-Strasse 2, 85748 Garching, Germany
\and
Department of Physics and Astronomy, University of British Columbia,
6224 Agricultural Road, Vancouver, British Columbia, 6658, Canada
}
\titlerunning{\textit{Planck}'s Dusty GEMS. IV. The Ruby}
\authorrunning{R. Ca\~nameras et al.} \date{Received / Accepted }

\abstract{We present an analysis of high-resolution ALMA
  interferometry of CO(4--3) line emission and dust continuum in the
  ``Ruby'' (PLCK$\_$G244.8$+$54.9), a bright, gravitationally lensed
  galaxy at z$=3.0$ discovered with the \textit{Planck} all-sky
  survey. The Ruby is the brightest of \textit{Planck}'s Dusty GEMS, a
  sample of 11 of the brightest gravitationally lensed high-redshift
  galaxies on the extragalactic sub-mm sky.  We resolve the
  high-surface-brightness continuum and CO line emission of the Ruby
  in several extended clumps along a partial, nearly circular Einstein
  ring with 1.4\arcsec\ diameter around a massive galaxy at
  $z=1.5$. Local star-formation intensities are up to 2000~M$_{\odot}$
  yr$^{-1}$ kpc$^{-2}$, amongst the highest observed at high redshift,
  and clearly in the range of maximal starbursts. Gas-mass surface
  densities are a few $\times 10^4$~M$_{\odot}$ pc$^{-2}$. The Ruby
  lies at, and in part even above, the starburst sequence in the
  Schmidt-Kennicutt diagram, and at the limit expected for star
  formation that is self-regulated through the kinetic energy
  injection from radiation pressure, stellar winds, and supernovae. We
  show that these processes can also inject sufficient kinetic energy
  and momentum into the gas to explain the turbulent line widths,
  which are consistent with marginally gravitationally bound molecular
  clouds embedded in a critically Toomre-stable disk.  The
  star-formation efficiency is in the range 1--10\% per free-fall
  time, consistent with the notion that the pressure balance that sets
  the local star-formation law in the Milky Way may well be universal
  out to the highest star-formation intensities. AGN feedback is not
  necessary to regulate the star formation in the Ruby, in agreement
  with the absence of a bright AGN component in the infrared and radio
  regimes.}

\keywords{galaxies: high redshift -- galaxies: evolution -- galaxies:
  star formation -- galaxies: ISM -- infrared: galaxies -- submillimeter:
  galaxies}

\maketitle
\section{Introduction}
\label{sec:introduction}

Vigorous star formation in high-redshift galaxies occurred in
environments with higher gas and stellar mass surface densities,
higher gas fractions, and strong turbulence compared to nearby
galaxies, with no obvious local kin. \citet{elmegreen99} already recognized
that massive, dense galaxy bulges must form most of their stars in one
to a few dynamical times (``maximal starburst''), in accordance with
the presence of a universal upper threshold of stellar mass surface
density \citep[][]{hopkins10}. How this limit is set is still a matter
of active debate; alternatives are either related to gas fragmentation
on kpc scales, or to the local kinetic energy injection from star
formation through radiation pressure or thermalized supernova ejecta,
and perhaps active galactic nuclei (AGN). \citet{andrews11} argued that
radiation pressure could explain the upper envelope of the
far-infrared luminosity of galaxies in the Schmidt-Kennicutt diagram
with a dependence on optical depth, and \citet{riechers13}
characterized such a maximal burst in an exceptional dusty starburst
at $z\sim6$.

Observers have stressed that rotation seems to be able to maintain the
gas marginally Toomre-stable \citep[e.g.,][]{nmfs09}, but
centrifugal support cannot be dominant on scales smaller than approximately
100~pc \citep[][]{toomre64, escala08}. Studying the local environments
and regulation mechanisms in the most intensely star-forming galaxies
at high redshift on small scales is particularly interesting in this
regard, and is now becoming feasible with ALMA for the most strongly
gravitationally lensed systems, such as SDP~81, for example
\citep[e.g.,][]{alma15,dye15,rybak15,hatsukade15,swinbank15,oteo16}.
Existing observations of the resolved Schmidt-Kennicutt law in
massive, dusty, gravitationally lensed high-redshift galaxies with
intense star formation and specific star-formation rates that fall
above those along the `main sequence' \citep[`starburst galaxies',
  e.g.,][]{elbaz11} suggest that star-formation intensities at a given
gas mass surface density are generally as high as, or factors of a few
greater than, those of galaxies near the high-redshift main sequence
\citep[][]{swinbank11, hatsukade15}. Investigating how high the
star-formation intensity (star-formation rate density) in such
galaxies can be is particularly interesting, because it allows us to
infer the mechanism that sets the upper boundary to how intensely
galaxies may form their stars.  Currently known high-redshift galaxies
generally fall factors of a few below the limit expected for galaxies
that form stars at rates near the Eddington limit as quantified by,
for example, \citet{andrews11}; however, suitable observations are still
very rare.

\begin{figure*}
\centering
\includegraphics[width=1.0\textwidth]{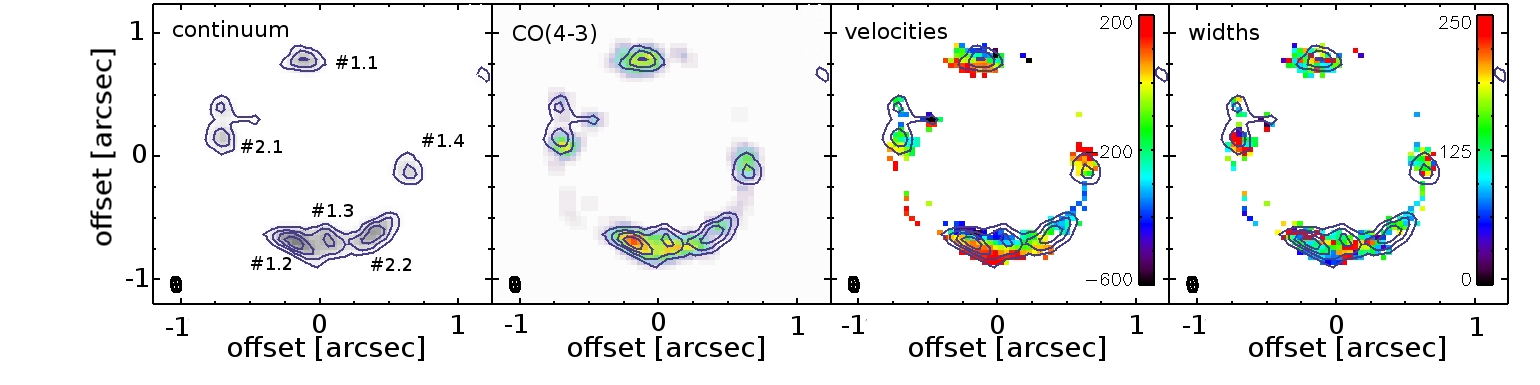}
\caption{Morphology and kinematics of CO(4--3) line emission in the Ruby. 
{\it Left to right:} Continuum and gas-mass surface brightness,
velocities (in km~s$^{-1}$) and Gaussian line widths
(FWHM$/2\sqrt{2\ln{2}}$, also in km~s$^{-1}$). Contours show the
continuum morphology at approximately 750~$\mu$m in the rest frame;
they start at 3$\sigma$, and increase in steps of 2$\sigma$. The
ellipse in the lower left corner of each panel shows the ALMA beam
size of $0.14\arcsec\times0.06\arcsec$.}
\label{fig:maps}
\end{figure*}

Here we report on ALMA 0.1\arcsec\ resolution observations from the
first long-baseline observing cycle of ALMA of the strongly lensed
$z=3$ galaxy PLCK~G244.8$+$54.9, the ``Ruby''. The Ruby is the
brightest source in our set of 11 Gravitationally Enhanced
subMillimeter Sources at redshifts $z=2.2-3.6$ (\textit{Planck}'s
Dusty GEMS), discovered with the \textit{Planck} all-sky survey.  The
modified blackbody fits to the \textit{Herschel}/SPIRE photometry of
\citet[][\citetalias{canameras15} hereafter]{canameras15} indicates an
observed peak flux density in the far-infrared of $\mu S_{\rm
  FIR}=1135\pm2$~mJy at 300 $\mu$m, which, when integrating over
8-1000~$\mu$m, corresponds to $\mu L_{\rm FIR} = 2.65 \pm 0.02 \times
10^{14}$~L$_{\odot}$ at $z=3.0$. SMA-850~$\mu$m and ALMA-3~mm
interferometry show a near-complete Einstein ring with
1.4\arcsec\ diameter, encircling a massive galaxy with a spectroscopic
redshift $z=1.52$, one of the highest-redshift lenses currently known
\citep[][\citetalias{canameras17} hereafter]{canameras17}, which itself
does however not contribute significantly to the bright far-infrared
emission. Detailed lens modeling with {\tt LENSTOOL} \citep[][]{jullo07}
suggests typical gravitational magnification factors between 10 and 40
for the different clumps along the ring (C17). An AGN may contribute
at most 10\% to the FIR luminosity, and the source falls near the
local far-infrared radio correlation \citepalias{canameras15}; AGN
feedback is therefore unlikely to play a major role in the gas
dynamics or regulation of star formation. Here we focus on the first
part of our ALMA program, high-resolution interferometry of the
CO(4--3) line with the long baseline array, which allows us to probe
the local star-formation law on scales below 100~pc, even at long
wavelengths of $\sim 3$~mm, which are best suited to probe relatively
low-J CO transitions. We also present a first discussion of how
feedback from star formation leaves its imprint on the resolved
star-formation law and kinematic properties of the gas.

The outline of the paper is as follows: In
Sect.~\ref{sec:observations} we present our observations and describe
the data reduction. In Sect.~\ref{sec:results} we describe the gas and
dust morphology of the Ruby and the kinematic properties of the
molecular gas as probed by the CO(4--3) line. We also investigate the
impact of differential lensing. In Sect.~\ref{ssec:regions12} we
discuss the intrinsic properties of the two independent regions within
the same galaxy, which give rise to the two systems of multiple,
gravitationally lensed images seen in the Ruby.  In
Sect.~\ref{sec:resolvedsflaw} we present the resolved
Schmidt-Kennicutt law between gas-mass surface density and
star-formation intensity. In Sect.~\ref{sec:limitsf} we discuss the
energy and momentum injection from star formation into the gas, and
investigate whether this can drive the gas turbulence, and regulate
star formation in the way suggested by the Schmidt-Kennicutt
diagram. We summarize our results in Sect.~\ref{sec:summary}.

Throughout the paper, we adopt the flat $\Lambda$CDM cosmology from
\citet{planck14xvi}, with $H_0$ = 68~km~s$^{−1}$ Mpc$^{−1}$,
$\Omega_{\rm m} = 0.31$, and $\Omega_{\rm \Lambda} = 1 - \Omega_{\rm
  m}$. At $z=3.005$ this implies a luminosity distance of 26.0~Gpc,
and a projected physical scale of 7.85~kpc~arcsec$^{-1}$. Where
appropriate, we explicitly mention the magnification factor $\mu$ to
mark results that have not been corrected for the gravitational
magnification, unless we refer to surface densities  
which are conserved by the lens.

\section{Observations and data reduction}
\label{sec:observations}

\subsection{ALMA band~3 interferometry of CO(4-3) and dust }
\label{ssec:almadata}

These ALMA Cycle 3 data (2015.1.01518.S, PI~Nesvadba) were taken on
October~23~2015 in the long baseline configuration C36-8 with maximum
baselines of over 10~km. The target was observed for 58~min with 35
antennas in excellent conditions with precipitable water vapor PWV of
around~0.6-1.0~mm and high phase stability (95-100~$\mu$m rms on
baselines of 6500~m). We had centered baseband 1 on the redshifted
CO(4--3) line at 114.888 GHz, and the remaining three spectral windows
onto the continuum at frequencies of 100.908~GHz, 102.783~GHz, and
112.897~GHz, respectively.

For the reduction of the ALMA data we used the standard manual scripts
with the Common Astronomy Software Application (CASA), applying
automatic and manual flagging of visibilities, calibrating bandpass,
phase and amplitude/flux, and using CLEAN to construct the
synthesized beam de-convolved images of the frequency data cubes and
continuum images. For antenna positions we used the best estimates
available from the entire campaign of baseline monitoring throughout
the long baseline observing in October and November 2015.

The data were imaged into cubes using the ``channel'' and ``velocity''
modes, ``briggs'' weighting with robust$=0.5$, and custom cleaning
masks using 1000 iterations. The rms in the final continuum image is
0.014 mJy beam$^{-1}$, and 0.34~mJy~beam$^{-1}$ in the CO(4--3) cube
for a spectral channel width of 24.4~MHz (63.7~km~s$^{-1}$). The beam
size is 0.14\arcsec$\times$0.06\arcsec\ along ${\rm PA}=55^\circ$.

\section{Results}
\label{sec:results}

\subsection{Continuum and gas morphology and gas kinematics}
\label{ssec:contresults}

\begin{table*}
\centering
\begin{tabular}{lccccc}
\hline
\hline
ID & RA       & Dec     & $\mu S_{dust}$   & $\mu^{1/2}$ R & $\mu$ \\
   & [J2000]  & [J2000] & [mJy]               & [arcsec] &       \\
\hline
1.1& 10:53:53.137 & +05:56:19.65 & 0.362$\pm$0.002 & 0.22& 22$\pm$2 \\
1.2& 10:53:53.145 & +05:56:18:18 & 0.569$\pm$0.002 & 0.23& 41$\pm$5 \\
1.3& 10:53:53.123 & +05:56:18.13 & 0.300$\pm$0.001 & 0.16& 30$\pm$3 \\
1.4& 10:53:53.090 & +05:56:18.84 & 0.220$\pm$0.002 & 0.15& 17$\pm$3 \\
\hline
2.1& 10:53:53.178& +05:56:18.97 & 0.403$\pm$0.002 & 0.19 & 11$\pm$2 \\
2.2& 10:53:53.101& +05:56:18.30 & 0.313$\pm$0.001 & 0.40 & 42$\pm$6 \\
\hline
\hline
\end{tabular}
\caption{Image-plane properties of individual counter images. Image 
ID as shown in Fig.~\ref{fig:maps}. Right ascension. Declination. 
Observed integrated continuum flux density at 3~mm. Circularized clump 
radius. Luminosity-weighted gravitational magnification factors from C17.}
\label{tab:clfind}
\end{table*}

In the left panel of Fig.~\ref{fig:maps} we show the continuum
morphology extracted from the two line-free windows 1 and 2 at 100.9
and 102.8~GHz as both gray-scale and contours. These frequencies
correspond to 404~GHz and 412~GHz in the rest-frame, respectively (and
to wavelengths of 742~$\mu$m and 728~$\mu$m, respectively).

The Ruby consists of several clumps belonging to two image systems
identified in C17, which form a partial Einstein ring around a massive
galaxy at $z=1.52$, with 1.4\arcsec\ diameter. The intervening lensing
source was not detected in our ALMA data cube down to the rms of our
continuum image. The continuum shows several clumps with maximal
surface brightness levels between (63$\pm$14)~$\mu$Jy~beam$^{-1}$ and
(130$\pm$14)~$\mu$Jy~beam$^{-1}$, that is, clumps are detected at
4.5$\sigma$ to 9$\sigma$. Individual clumps are spatially resolved, at
least along the major axis, with sizes between 0.25\arcsec\ and
0.4\arcsec, seen with a beam of 0.14\arcsec$\times$0.06\arcsec. The
projected sizes in the image plane, and maximal continuum surface
brightnesses of each clump are listed in Table~\ref{tab:clfind}.

We extrapolate the dust SED from \citetalias[][]{canameras15} to find
an expected source-integrated continuum flux density at the frequency
of our ALMA observations of 2.75~mJy. Integrating the continuum flux
over the entire ring of the Ruby, we recover 73\%\ of this flux
density, that is, 2.00$\pm0.05$ mJy (not correcting for lensing). Most of
the missing flux is likely to be in the extended diffuse regions along
the Einstein ring seen with the Submillimeter Array at 850~$\mu$m
(C17); these regions are not too extended to be seen with ALMA with
the long baselines that we used, but are too faint to be significantly
detected with our small beam size at 3~mm. In the following analysis
we focus on the properties of the intensely star-forming clumps,
so that the missing flux will not significantly affect our analysis.

We construct maps of the CO(4--3) line observed in spectral window 3 by
fitting single Gaussian profiles to each spatial pixel with a custom
IDL routine based on the MPFIT algorithm of \citet{markwardt09}.
Fig.~\ref{fig:maps} shows the resulting gas morphology, velocity maps,
and maps of Gaussian line widths, with the continuum
morphology overplotted as contours. The line emission is dominated by
several bright clumps, some of which are resolved into a few
resolution elements. Maximal CO line surface brightness levels are
between 200 and 600 mJy km s$^{-1}$ beam$^{-1}$, with typical
uncertainties of $20-40$~mJy km s$^{-1}$ beam$^{-1}$. Velocities measured 
locally in each pixel span a range between $-600$ and $200$ km s$^{-1}$ 
relative to z$=3.005$, with typical uncertainties of $30-40$ km s$^{-1}$. 
The velocity gradients in the northern and southern clumps are nearly
perpendicular to the magnification direction, which shows that we
resolve the source also along that direction. To our knowledge, this
is the first time that this has been achieved in the millimeter for
a gravitationally lensed image.

Velocity dispersions (Gaussian widths) are between 25~km~s$^{-1}$ and
200~km~s$^{-1}$, with typical uncertainties of $40-50$ km s$^{-1}$,
and with a few smaller regions between clumps having widths up to
approximately 330~km~s$^{-1}$. This could be a signature of blending of line
emission from multiple sources. For a sound speed $c_{\rm
  s}=\sqrt{\gamma\ k_{\rm B}\ T / m_{\rm H2}}=0.6$ km s$^{-1}$, with
Boltzmann constant $k_{\rm B}$, temperature T$=50$~K
\citepalias[][]{canameras15}, and molecular mass of $H_{\rm
  2}=3.24\times 10^{-24}$ g, these widths correspond to highly
supersonic velocities with Mach numbers between ${\cal M}=40$ and 550.

\subsection{Morphological and kinematic substructure of the Ruby}
\label{ssec:clumpresults}

The morphology and kinematics of the Ruby are complex. C17 identify
multiple lensed images of two physically distinct regions, which
partially overlap. In order to examine the morphological and kinematic
substructure in the Ruby in a robust, reproducible way, we analyzed the
CO(4--3) data cube with the Clumpfind algorithm
\citep{williams94}. Clumpfind identifies contiguous regions of line
emission in three-dimensional data cubes taking all spatial and
velocity information into account, and returns the position, size,
aspect ratio, and integrated emission line properties of each clump.

We ran Clumpfind in steps of $2\sigma$, starting at $2\sigma$, as
suggested by \citet{williams94}, where $\sigma=2$~mJy is the root mean
square of the CO(4--3) data set measured from a cube that was smoothed
by three~pixels along the two spatial axes, and five~pixels along the
spectral axis. This is comparable to the size of the beam and the line
width, respectively, and hence does not lead to a loss in spatial
resolution while marginalizing over the details of the line profile
per pixel, and hence maximizing the signal-to-noise along the spectral
axis. 

With Clumpfind we identified a total of 12 clumps at $\ge 5
\sigma$. We also inverted the cube, to provide another constraint
  on the potential number of spurious clumps, finding none with
  $>4\sigma$ significance. Visual inspection of these clumps in the
CO data cube showed that 11 of them are extended over at least the
size of the beam, and their peak is detected at significances of up to
14~$\sigma$. All can be associated with the multiple images identified
visually by C17 from the peaks of the flux map and orientation of the
velocity gradients. However, blueshifted and redshifted gas in a given
counter image was typically identified as two individual clumps. The
12th and faintest clump appears to be spurious, and was discarded from
the subsequent analysis.

\begin{figure}
\centering
\includegraphics[width=0.5\textwidth]{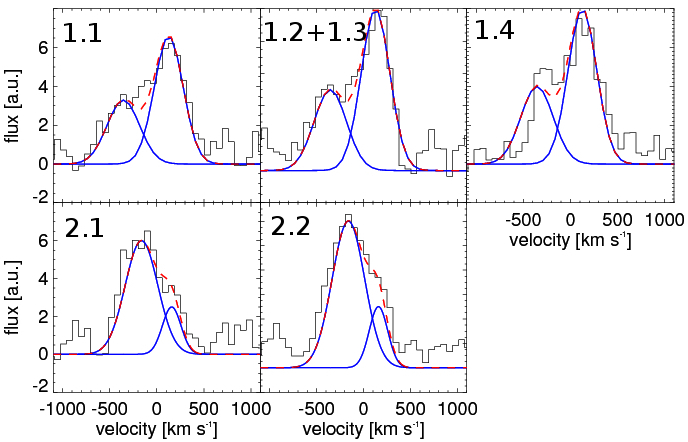}
\caption{Spectra of individual counter images as assembled from clouds
  identified with Clumpfind, as labeled in
  Fig.~\ref{fig:maps}. Images \#~1.2 and \#~1.3 are blended and
  identified as a single clump by Clumpfind. The dark blue and red
  lines show fits with Gaussian functions to the total line
  profile of each image system shown in Fig.~\ref{fig:familyspec},
  scaled to the total flux in each image. The mismatch between the
  blue component in image \#~1.4 can be explained with partial overlap
  with image \#~2.2.}
\label{fig:familyspec}
\end{figure}

\begin{figure}
\centering
\includegraphics[width=0.4\textwidth]{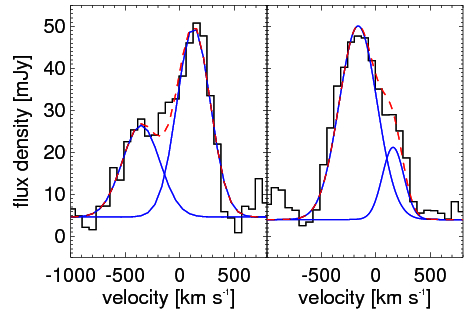}
\caption{Integrated spectra of image system 1 (left) and 2 (right), 
  summed over images \#~1.1 and \#~1.2, and \#~2.1, and \#~2.2,
  respectively. Solid blue lines show individual Gaussian components
  of the line profiles, while the red dashed line shows the sum of
  both components.}
\label{fig:imagespec}
\end{figure}

To provide an independent verification of the previous identification
of image systems by C17 we reassembled clumps into individual lens
images, which are labeled in the left panel of Fig.~\ref{fig:maps}.
Figure~\ref{fig:familyspec} shows the resulting spectra integrated
over all images. This would not be the case, if, for example, 
Clumpfind had missed significant parts of the flux. Images~\#1.2 and~\#1.3 are
strongly blended, and were not identified as separate clumps by
Clumpfind. Discrepancies between the line profile of image \#1.4 and
the other images of system~1 can be explained with partial overlap of
images \#1.3 and \#2.2, or potentially different magnification factors
for the blueshifted and redshifted gas. Note that the blue and red
lines in Fig.~\ref{fig:familyspec} are not fits of Gaussian
functions to the spectra of individual counter images, but rescaled
versions of the sum of all counter images in a given image system
shown in Fig.~\ref{fig:maps}, keeping the same line width and redshift
as in the total image of each system, and also the same ratio between
the two line components. The fluxes of these spectra are listed in
Table~\ref{tab:imagespec}. The similarity of these line profiles
confirms the previous assignment of individual images to image
systems, which C17 did based on the lens modeling. The combined
spectrum of each image system is shown in Fig.~\ref{fig:imagespec}.

\section{Intrinsic properties of regions 1 and 2}
\label{ssec:regions12}

We will now focus on two of the brightest, most isolated images, \#1.1
and \#2.1, to analyze the properties of the two regions in the source
plane that are magnified by the gravitational lens in front of the
Ruby. Since all other images are either from system~1 or 2, they would
provide the same physical constraints at lower signal-to-noise
ratio. Images \#1.2\ and \#2.2\ are somewhat brighter, but are also
contaminated with partially overlapping adjacent images, and their
morphology and gas kinematics is therefore more difficult to
interpret. Images \#1.1\ and \#2.1\ are also least affected by
systematic uncertainties from the lens reconstruction.

\begin{figure*}
\centering
\includegraphics[height=0.22\textheight]{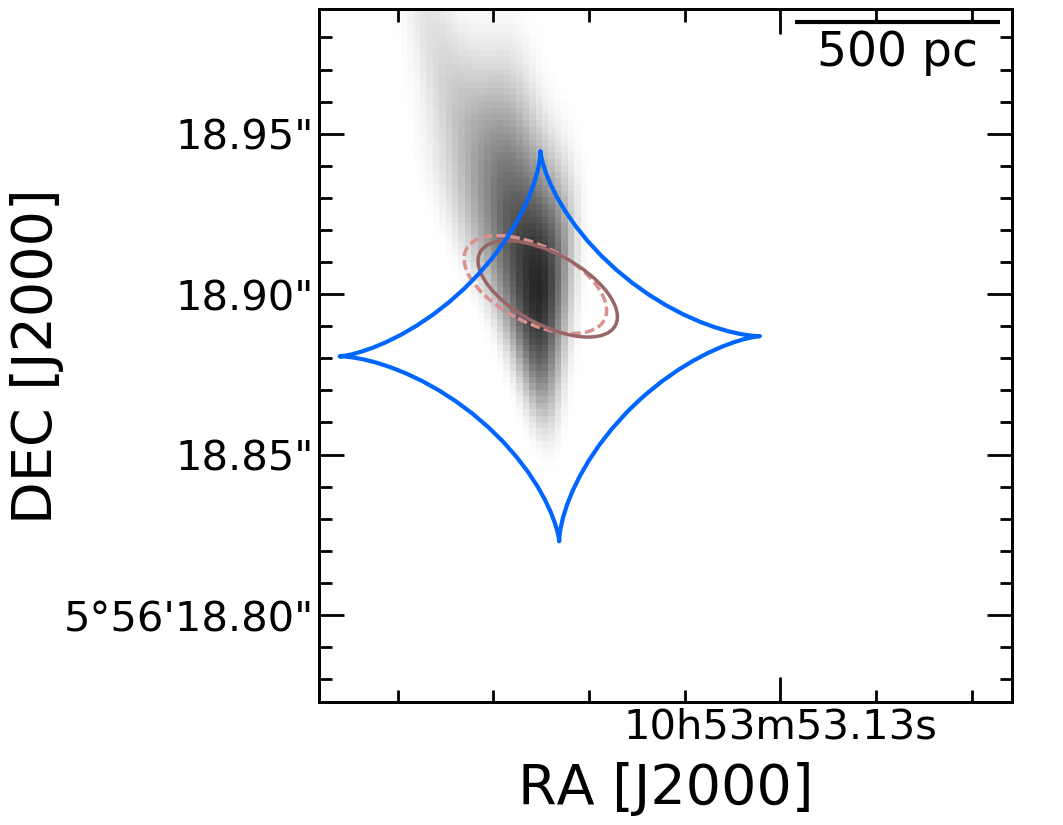}
\includegraphics[height=0.22\textheight]{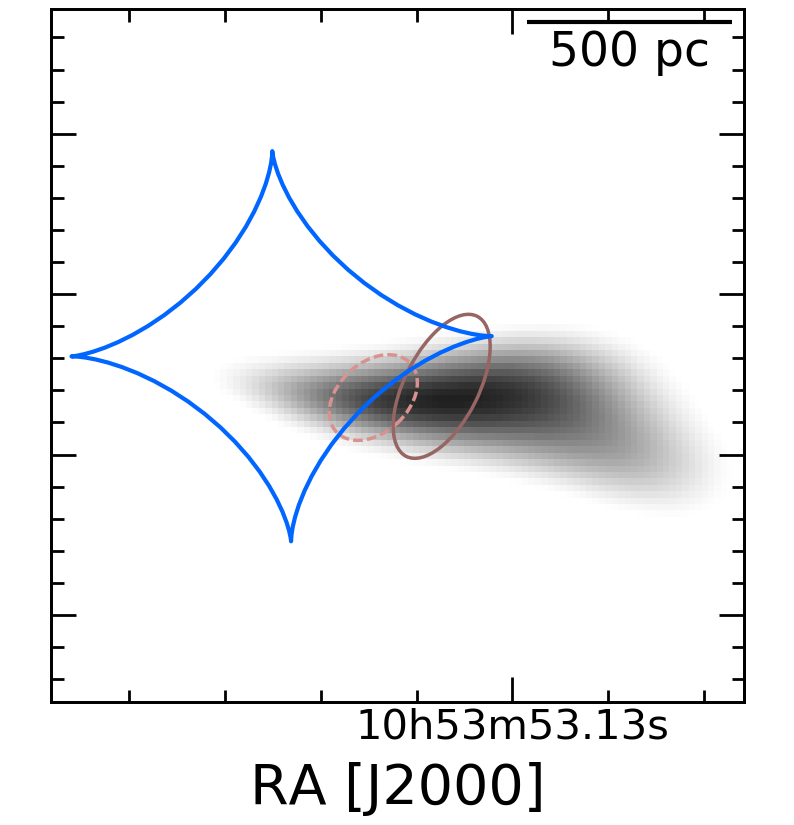}
\caption{Reconstructed source-plane morphology of the integrated and
    beam-convolved CO(4--3) line image of regions~1 {\it (left)}
  and 2 {\it (right)} as obtained with {\tt LENSTOOL}. Both regions are
  spatially well resolved by our ALMA data. Blue diamond-shaped lines
  indicate the position of the internal caustic line at $z=3.005$. The
  solid and dotted brown lines show the 3-sigma positional uncertainty
  on the luminosity-weighted centroid of each region, for the line and
  continuum emission, respectively.}
\label{fig:sourceplane}
\end{figure*}

The source-plane reconstruction of C17 obtained with {\tt LENSTOOL} suggests
that these two image systems trace two independent regions that are
separated by 470~pc in the source plane (Fig.~\ref{fig:sourceplane}). 
Region~1 samples a region with intrinsic maximal side lengths in the 
source plane of approximately 1.4~kpc and 0.7~kpc along the major and 
minor axis, respectively, while these lengths in region~2 amount to 
1.2~kpc and 0.3~kpc along the major and minor axis, respectively. Both 
regions are well spatially resolved in our ALMA data, and likely extend 
beyond the part of the galaxy which is most strongly magnified by the 
gravitational lens. The sizes were measured from the reconstructed 
source-plane morphology determined through the CleanLens algorithm within 
{\tt LENSTOOL} \citep[][]{sharon12}, and correspond to 3$\sigma$ isophotal 
sizes in the image plane. Corresponding areas are 0.5~kpc$^2$ and 0.3~kpc$^2$,
respectively, and were derived by summing over the area of all pixels
in the reconstructed source-plane image (Fig.~\ref{fig:sourceplane}). 
Taking into account their gravitational magnification factor and the beam 
shape, we reach maximal spatial resolutions of 64~pc and 160~pc along 
the caustic line in region~1 and 2, respectively.

\begin{table}
\centering
\begin{tabular}{lcccccc}
\hline
\hline
ID & Redshift & v & FWHM & $\mu I_{CO}$  \\
   & & [km s$^{-1}$]  & [km s$^{-1}$] &[Jy km s$^{-1}$] \\
\hline
  1.1 & 3.00373$\pm$0.00008  & -381$\pm$25  & 374$\pm$58  & 5.6$\pm$1.1  \\
      & 3.00539$\pm$0.00004  & 117$\pm$11   & 436$\pm$27  & 14.7$\pm$1.2 \\
  1.2 & 3.00371$\pm$0.00005  & -387$\pm$14  & 562$\pm$34  & 19.9$\pm$1.6 \\
      & 3.00548$\pm$0.00002  & 145$\pm$6    & 358$\pm$15  & 22.3$\pm$1.3 \\
  1.3 & 3.00400$\pm$0.00004  & -300.5$\pm$11& 449$\pm$26  & 13.8$\pm$1.1 \\
      & 3.00541$\pm$0.00009  & 124$\pm$28   & 495$\pm$68  & 6.4$\pm$1.2  \\
  1.4 & 3.00448$\pm$0.00021  & -155$\pm$63  & 414$\pm$161 & 2.8$\pm$1.4  \\
      & 3.00561$\pm$0.00005  & 184.2$\pm$16 & 345$\pm$38  & 7.2$\pm$1.0  \\
\hline
Sum   & 3.00382$\pm$0.00004  & -355$\pm$13  & 406$\pm$31  & 9.5$\pm$1.0  \\
      & 3.00542$\pm$0.00002  & 127$\pm$6    & 359$\pm$14  & 17.3$\pm$0.9 \\
\hline
  2.1 & 3.00441$\pm$0.00004  & -178$\pm$12  & 525$\pm$28  & 17.9$\pm$1.2  \\
      & 3.00534$\pm$0.00011  & 101$\pm$34   & 1030$\pm$81 & 17.0$\pm$1.8  \\
  2.2 & 3.00451$\pm$0.00003  & -147$\pm$10  & 399$\pm$24  & 21.5$\pm$1.7  \\
      & 3.00556$\pm$0.00007  & 169$\pm$20   & 173$\pm$50  & 3.6$\pm$1.3   \\
\hline
Sum & 3.00446$\pm$0.00002 & -162$\pm$7     & 409$\pm$16  & 20.1$\pm$1.0  \\
    & 3.00553$\pm$0.00005 &  158$\pm$14    & 233$\pm$34  & 4.3$\pm$0.8   \\
\hline
\hline
\end{tabular}
\caption{Redshifts, centroid velocities relative to $z=3.005$, FWHM
  line widths, and integrated line fluxes in individual counter images
  of the Ruby. We also provide fit results for the sum of multiple
  images from each system as shown in Fig.~\ref{fig:familyspec}.}
\label{tab:imagespec}
\end{table}

\subsection{Dynamical mass and mass surface densities}

The angular separation of 470~pc between these two regions is smaller
than the size of each region, which suggests that both are within the
gravitational potential of the same galaxy or an advanced galaxy
merger near coalescence. This is also suggested by the excellent
agreement in velocity between the two regions, which are within
200~km~s$^{-1}$ of one another. The velocity difference between these
two regions (measured from the peak of the Gaussian line profiles in
their integrated spectra) is less than the range of velocities,
$\Delta v$, which we measure in the velocity maps of each individual
region, and which are given relative to a single reference redshift of
$z=3.005$. In region~1, we find a total velocity range of $\Delta
v_1=550\pm35$~km~s$^{-1}$, and in region~2, this range is $\Delta v_2=
260\pm27$~km~s$^{-1}$. Integrated luminosity-weighted Gaussian line
widths, $\sigma={\rm FWHM}/2\sqrt{\ln{2}}$, in each clump are between
approximately $130\pm38$ km s$^{-1}$ and $330\pm45$~km s$^{-1}$.

We measured minimal and maximal velocities directly from the map,
without the use of a model. Commonly used algorithms, such as, for example,
kinemetry \citep[][]{krajnovic06} rely on fitting up to 15 free
parameters internally, which we cannot provide with our data set,
which is resolved to only $2-4$ independent resolution elements along
the kinematic major axis.

If these gradients approximately sample gravitational motion within a
rotating disk, we can provide dynamical mass estimates by setting
$M_{dyn}= v^2 R / G$. Here $v$ is the circular velocity, which we
approximate as $\Delta v / 2\ \sin{i}$, $R$ the radius, and $G$ the
gravitational constant. The small spatial sizes we probe in the source
plane make it unlikely that the gas kinematics probe two fairly
isolated, approximately virialized galaxies in the early stage of a
merger. Using the above measurements, and with intrinsic sizes
$R_1 = 700$~pc and $R_2 =600$~pc for regions 1 and 2, we find
dynamical mass estimates of ($1.2\pm0.3)$ and $(0.24\pm0.05) \times
10^{10} \sin{i}^{-2}$ M$_{\odot}$, respectively, and corresponding mass
surface densities of $(8\pm1.7) \times 10^{9} \sin{i}^{-2}
$~M$_{\odot}$~kpc$^{-2}$ and $(2\pm0.4) \times 10^{9} \sin{i}^{-2}$
M$_{\odot}$~kpc$^{-2}$. We calculated mass surface densities by
dividing the above dynamical mass estimates by the surface of circular
regions of radius $R_1$ and $R_2$, respectively. Note that these
estimates are not corrected for inclination effects and potential
offsets between rotation and magnification direction; together, these
typically lower the observed relative to the intrinsic mass by factors
of $2-4$, which would imply intrinsic dynamical masses of regions 1
and 2 of approximately $(3.6\pm0.8)$ and $(0.8\pm0.2) \times 10^{10}$
M$_{\odot}$, respectively, and mass surface density estimates of
$(3.7\pm0.8)$ and $(0.6\pm0.1) \times 10^{10}$~M$_{\odot}$~kpc$^{-2}$,
respectively. The greater of these values is close to the highest
stellar mass surface densities observed in massive early-type galaxies
and globular clusters at low redshift \citep[e.g.,][]{hopkins10},
$\sim 5-10\times 10^{10}$ M$_{\odot}$ kpc$^{-2}$. \citet{kauffmann03}
argued that high mass surface densities are a signature predominantly
of very massive galaxies, so that this finding adds another piece to
the considerable body of evidence of the evolutionary link between
massive low-z elliptical galaxies and the most intense starbursts at
redshifts $2-3$, \citep[e.g.,][]{blain02,tecza04,swinbank06,amblard11}.

\subsection{Global gas stability and fragmentation scale}

Similar to galaxies in the field, the disk in the Ruby appears to be
overall Toomre-stable.  A number of parametrizations of the Toomre
parameter, $Q$, have been proposed in the literature, depending on gas
configuration, and whether stellar or gas mass surface density
dominates. Here we use the parametrization of \citet{genzel14} to set
$Q=a\ v_{\rm c}\ \sigma / \pi\ R\ G\ \Sigma_{\rm gas}$, with $a=2$ as
is appropriate for the monotonically rising part of a rotation
curve. $v_{\rm c}$ is the rotational velocity, $\sigma_{\rm T}$ the
turbulent velocity dispersion. $R$ is the disk radius (we adopt the
700~pc in \# 1.1, i.e., the largest radius magnified by the lens) and
$G$ is the gravitational constant. For the gas-mass surface density,
$\Sigma_{\rm gas}$, we use a typical value for the Ruby of $1\times
10^{4}$~M$_{\odot}$ pc$^{-2}$ (Sect.~\ref{ssec:schmidtkennicutt}).
This gives $Q=1.0\pm0.3$. Hence the gas in the Ruby appears to be
globally marginally Toomre-stable.

Clumpfind identifies only a single contiguous component associated
with image \#1.1, with a FWHM size of 0.2\arcsec$\times$0.12\arcsec,
corresponding to an (averaged) diameter of 0.16\arcsec. In image
\#2.1, the algorithm identifies four individual clumps with sizes
between 0.1\arcsec\ and 0.2\arcsec. Corrected for the
luminosity-weighted average magnifications, this corresponds to
source-plane diameters between approximately 40~pc and 200~pc in the two
images at $z=3.005$. These sizes are comparable to or smaller than the
sizes of individual small clumps identified in unlensed dusty
starburst galaxies at similar redshifts, which have typical diameters
of 200~pc \citep[][]{iono16}.

This range also corresponds to the range of sizes expected for gas
clouds in a fragmenting disk. Expected fragmentation scales, $L_{\rm
  J}$, expected from a classical Jeans analysis at these mass surface
densities scale as $L_{\rm J} = \sigma_{\rm t}^2 /
2\ \pi\ G\ \Sigma_{\rm gas}$, where $\sigma_{\rm t}$ is the velocity
dispersion of the gas, $G$ the gravitational constant, and
$\Sigma_{\rm gas}$ the gas-mass surface density. The corresponding
mass for a uniform-density sphere, $M_{\rm J}$, can be written as
$M_{\rm J} = \pi\ \sigma_{\rm t}^2\ \Sigma_{\rm gas}$. For Gaussian
widths between 100 and 170~km s$^{-1}$ and mass surface densities of
approximately $10^4$ M$_{\odot}$ pc$^{-2}$, we find cloud sizes of 
approximately $40-100$~pc, and masses between $1$ and $3 \times 10^8$
M$_{\odot}$. These are smaller than the gas masses we see in
individual regions of the Ruby. Cloud overlaps or a larger
fragmentation scale in clumpy high-z galaxies than suggested by this
simple analysis \citep[][]{romeo14} might cause this difference.

\subsection{A note on differential lensing}

Depending on the relative morphology of dust and gas, differential
lensing may produce major biases in studies of strongly lensed
galaxies at low spatial resolution \citep[e.g.,][]{serjeant12}. 
However, these estimates were made on toy models, which is unavoidable 
given our limited knowledge of the internal structure of dusty 
high-redshift starburst galaxies. The exceptionally high spatial 
resolution of our ALMA data allows us to quantify the potential biases 
from differential lensing in a direct way.

\begin{figure}
\centering
\includegraphics[width=0.45\textwidth]{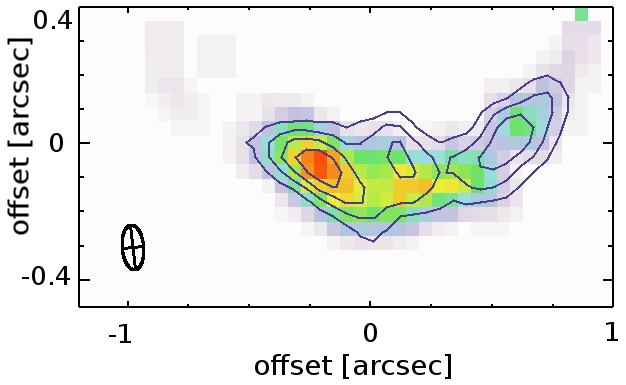}
\caption{Zoom onto the Southern part of the Ruby, showing the CO(4--3)
  morphology as color image, and the dust morphology as contours. We note
  the small offset between the emission-line and continuum peak.}
\label{fig:offsets}
\end{figure}

We do find small positional offsets between the peaks of the dust and
the CO line emission (Fig.~\ref{fig:maps} \& \ref{fig:offsets}) of up to 0.2\arcsec\
  in two clumps seen in the image plane, which correspond to an
  intrinsic offset in the source plane (Fig.~\ref{fig:sourceplane}). For
the other clumps, we do not find a measurable offset between the
intrinsic dust and gas morphology in the source plane. Since dust and
gas morphologies were derived with the same observations and are at
almost the same frequencies, these offsets must be intrinsic, and show
that we are resolving the internal structure of the star-forming
regions in the Ruby.

We use luminosity-weighted maps of CO and continuum emission to infer
the impact of the differences in gravitational magnification on the
total estimates of the gas and dust luminosity. When summing over all
pixels that are seen in both gas and dust, we find luminosity-weighted 
average magnification factors of $33.8\pm6.4$ and $43.6\pm8.2$ for the 
FIR continuum luminosity and CO(4--3) line luminosity, respectively. 
This corresponds to a difference of approximately 22\% for the integrated 
values, and is not greater than other systematic and measurement 
uncertainties in this kind of study.

For image \#1.1, a similar analysis gives luminosity-weighted
average magnification factors of 21.7$\pm$4.2 and 19.8$\pm$3.3 for
the continuum and line luminosities, respectively, while for \#2.1
we find factors of 10.7$\pm$2.6 and 9.0$\pm$2.2, respectively. This
corresponds to approximately 10\% for image \#1.1, and 20\% for image
\#2.1. Again, these differences are not very large compared to other
systematic and measurement uncertainties, and reflect the intrinsic
offsets between gas and dust on the small spatial scales on which we
resolve the Ruby in the source plane.

\section{Resolved star-formation law and self regulation}
\label{sec:resolvedsflaw} 

\subsection{Schmidt-Kennicutt law}
\label{ssec:schmidtkennicutt}

\begin{figure*}
\centering
\includegraphics[width=0.49\textwidth]{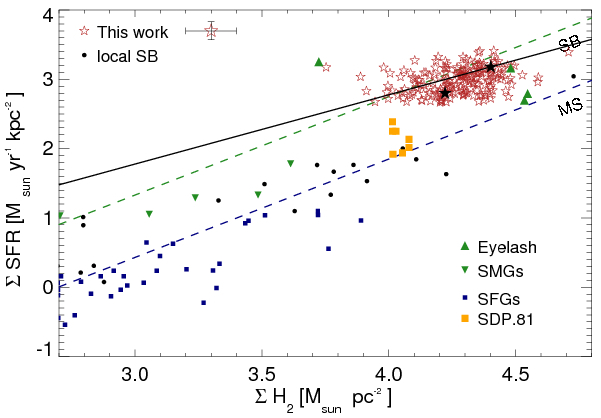}
\includegraphics[width=0.49\textwidth]{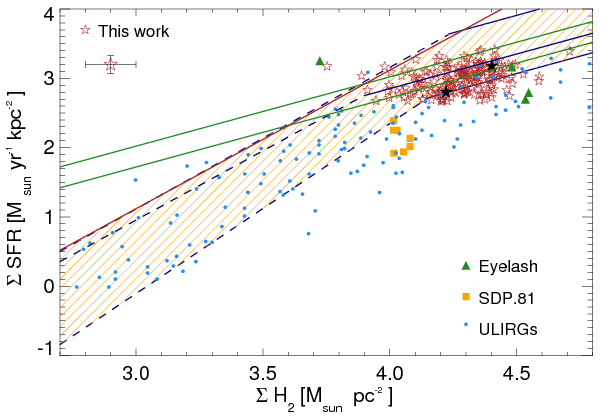}
\caption{{\it Left:} spatially resolved Schmidt-Kennicutt law in
  PLCK~G244.8$+$54.9 (red stars) and best fit relation (solid black line). 
  The lower and upper black star
  indicate the luminosity-weighted average position of images~\#2.1
  and \#1.1, respectively. Other symbols indicate: submillimeter
  galaxies at $z\sim2$ \citep[green upside-down triangles,][]{bothwell10}, 
  local starbursts \citep[black circles,][]{kennicutt98},
   spatially-resolved data from the Eyelash
  \citep[dark green triangles,][]{swinbank11} and SDP.81 \citep[yellow
    squares,][]{hatsukade15}. The green and dark blue dashed lines
  indicate the Schmidt-Kennicutt relationship for high-z starburst
  galaxies and main sequence galaxies of \citet{daddi10},
  respectively. All star-formation rates are given for a Chabrier
  initial mass function, and are a factor~1.8 lower than in many other
  figures showing the Schmidt-Kennicutt relationship (see text for
  details).  {\it Right:} same Schmidt-Kennicutt diagram, showing
  the position of maximum starbursts in several models, and the
  position of the Ruby relative to them. Light blue circles indicate
  nearby Eddington-limited starbursts \citep[][]{andrews11}. The dark
  green solid lines show the mechanical input from stellar winds and
  supernova remnants for 100\%\ and 50\%\ efficiency,
  respectively. The dark blue lines and yellow hatched region show the
  models of \citet{andrews11} for an Eddington-limited starburst with
  a range of opacities between 10~and~30~cm$^2$ g$^{-1}$. Solid lines
  show the optically thick limit above the critical gas-mass surface
  density for the optically thick case, dashed lines the optically
  thin case below that limit. The solid blue line shows the
  relationship of \citet{ostriker11}, also for optically thick and
  optically thin gas. The red line shows the model of
  \citet{faucher13}.}
\label{fig:sklaw}
\end{figure*}

Star-formation intensity (projected star-formation rate density) is
closely related to the local gas-mass surface density in galaxies over
more than six orders of magnitude in star-formation intensity, and
this relationship seems to be linear for gas-mass surface densities
above approximately 10~M$_{\odot}$ pc$^{-2}$ \citep[][]{bigiel08}. The zero
point of this relationship is a measure of the efficiency with which
galaxies turn their gas into stars. Most studies of star-forming
regions in the Milky Way and nearby galaxies suggest this efficiency
is around 1\% per gas free-fall time \citep[e.g.,][]{krumholz07}. The
reasons for this are not yet fully understood, although there is a
growing consensus in the literature that the turbulence within
molecular clouds and filaments plays a major role in establishing this
parameter \citep[][]{klessen00, krumholz05, li05, hennebelle07,
  bournaud10, audit10}.

At high redshift, it has been suggested that galaxies falling onto the 
main sequence of star-forming galaxies could have a lower
star-formation efficiency than starburst galaxies that fall above this
relationship \citep[e.g.,][]{daddi10, genzel10}, perhaps a consequence
of the higher pressures reached in the dense gas that is filling the
deep gravitational potential wells of the most intensely star-forming
high-redshift galaxies \citep[][]{swinbank11,swinbank15}. Likewise,
the energy injection from intense star formation should limit the
maximal rate with which gas can be turned into stars
\citep[e.g.,][]{heckman90, lehnert96a, lehnert96b, veilleux05,
  andrews11, murray05}.

The Schmidt-Kennicutt diagram plotted in the left panel of
Fig.~\ref{fig:sklaw} shows the local star-formation intensities as a
function of the local gas mass surface densities extracted from each
spatial pixel \citep[][]{kennicutt89}. To estimate the star-formation
rates, we measured the average continuum flux density in each
  0.04\arcsec$\times$0.04\arcsec\ pixel in our line-free ALMA spectral
windows 1 and 2, and extrapolated along a modified black body curve
with a global temperature, $T=50$~K, and $\beta=2.0$ \citep[as used by]
[for the Ruby]{canameras15}, to translate these monochromatic flux
densities into FIR luminosities.

We follow \citet{kennicutt89} in estimating the star-formation rates
by setting ${\rm SFR}=4.5\times 10^{-44} L_{\rm FIR}$, where SFR is
given in M$_{\odot}$ yr$^{-1}$. $L_{\rm FIR}$ is in erg s$^{-1}$,
integrated between 8~$\mu$m and 1000~$\mu$m in the rest
frame. Resulting star-formation rates are then corrected by a
factor~1.8 downward to adopt the now more commonly used Chabrier
initial mass function instead of the Salpeter initial mass function
originally adopted by \citet{kennicutt89}. \citet{tacconi08} showed
that Salpeter initial mass functions would lead to stellar mass
estimates for high redshift sub-millimeter galaxies that are
inconsistent with gas and dynamical mass estimates.

This estimation procedure results in local star-formation intensities
of $220-2200$~M$_{\odot}$ yr$^{-1}$ kpc$^{-2}$, which is higher by
approximately two orders of magnitude than in low-redshift starburst galaxies
\citep[e.g.,][]{kennicutt98}. It is also higher by factors of a few
than in typical high-redshift galaxies that are closer to the main
sequence. For example, star-formation intensities in H-ATLAS
J$09011.6+003906$ (SDP.81) are below 200 M$_{\odot}$ yr$^{-1}$
kpc$^{-2}$, as found by \citet{rybak15} with an analysis that was also
based on star-formation rates measured pixel by pixel.

Such values are akin to those measured in the brightest, not
gravitationally lensed dusty starburst galaxies at high redshifts, for
which measurements of the dust morphology have been obtained at
resolutions $\la0.5$\arcsec\ \citep[][]{ikarashi15, iono16}. Total
sizes of the most actively star-forming regions in these galaxies are
approximately $1-4$~kpc \citep[][]{ikarashi15, iono16}, compared to major axis
sizes of approximately 1.4~kpc and 1.2~kpc, respectively, for regions~1 and~2
seen in the Ruby (Sect.~\ref{ssec:regions12}). We can therefore assume
that the star-formation properties as probed in the Ruby are
representative of the most intense high-redshift starbursts.
\citet{oteo16} recently reported comparable star-formation intensities
of up to 3000~M$_{\odot}$ yr$^{-1}$ kpc$^{-2}$, over scales of approximately
200~pc, in a pair of strongly lensed galaxies at $z=3.4$ discovered in
ALMA calibration data, but also point out that parts of the FIR
continuum could potentially come from an AGN, which is not the case
here \citepalias{canameras15}.

The abscissa of Fig.~\ref{fig:sklaw} shows the molecular gas mass
surface density, which we derived from the CO(4--3) surface
brightness. The linearity of the relationship between CO and FIR
luminosity of high-z galaxies suggests that the brightness of the
J$=4-3$ transition is dominated by gas-mass surface density, not gas
excitation \citep[][]{greve14}. We follow \citet{solomon97} to
estimate molecular gas masses from the CO line flux of each spatial pixel, 
and adopt a CO-to-H$_2$ conversion factor of
$\alpha_{\rm CO}=0.8$~M$_{\odot}$ [K km s$^{-1}$ pc$^2$]$^{-1}$, which is a
commonly adopted value in studies of dusty, intensely star-forming
high-redshift galaxies \citep[e.g.,][]{tacconi08, bothwell10,
  hatsukade15}, and also agrees with the global dust-to-gas ratio
estimates of \citetalias{canameras15} from the FIR-to-millimeter dust
photometry of the GEMS. Less intensely star-forming high-redshift
galaxies appear to be better characterized by a factor that is closer
to the standard conversion factor found in the Milky Way, which is
approximately five~times greater \citep[see][for a recent review]{bolatto13},
but any correction by more than a factor of approximately two would lead to gas
masses and gas mass surface densities in the Ruby that are greater
than the dynamical mass estimate and derived surface densities, which
would therefore be unphysical.

A correction factor for which no consensus has yet been reached is the
flux ratio between mid-J CO lines and CO(1-0), for which the
Schmidt-Kennicutt law has initially been calibrated. For CO(4--3),
\citet{carilli13} and \citet{bothwell13} found
$r_{43/10}=L^\prime_{4-3}/L^\prime_{1-0}\sim0.4$ in high redshift
starburst galaxies, however, these have on average much lower
star-formation intensities. Galaxy-integrated measurements can also be
contaminated with CO(1-0) emission from diffuse gas not associated
with star-forming clouds \citep[][]{ivison10}, thereby preferentially
increasing the apparent CO(1-0) flux, so that estimates of molecular
gas masses associated with star formation from mid-J CO lines might be
more reliable than those from CO(1-0). This could happen to a
different degree in unlensed galaxies in the field compared to the
Ruby, which samples a small, intensely star-forming region.

In the following, we therefore adopt r$_{43/10}=0.6$, which
corresponds to the measured value in the Eyelash
\citep[][]{danielson11} and a stack of lensed galaxies from the South
Pole Telescope sample \citep[][]{spilker14} and falls between the
values adopted by \citet{tacconi08} and \citet{carilli13}. 

With these assumptions and corrections, resulting gas-mass surface
densities are within $10^{3.7-4.7}$ M$_{\odot}$ pc$^{-2}$, and most
apertures fall between $10^{4.1}$ and $10^{4.5}$ M$_{\odot}$
pc$^{-2}$. The Ruby falls therefore well above the locus of
main sequence galaxies at high redshifts in the left panel of
Fig.~\ref{fig:sklaw}, and for some apertures even above the typical
sequence of intense starbursts found from galaxy-integrated
measurements \citep[][]{daddi10}, which could at least partially be
due to the lower physical resolution in observations of unlensed
galaxies.

\subsection{Self regulation in a maximum starburst}
\label{ssec:selfreg}

The high star-formation intensities we find in the Ruby are consistent
with the extreme values of $\ga 1000-2000$ M$_{\odot}$ yr$^{-1}$
kpc$^{-2}$ expected for galaxies that form the bulk of their stars
within one or a few crossing times \citep[``maximal
  starbursts''][]{elmegreen99, tacconi06, riechers13}. They are also
greater by factors of a few than the intensities measured previously
in strongly gravitationally lensed dusty starburst galaxies at high
redshift (Fig.~\ref{fig:sklaw}), meaning that our observations of the Ruby
allow us to push even further towards probing the most intensely
star-forming systems.

Several theoretical studies have proposed that galaxies in this part
of the Schmidt-Kennicutt diagram should be self-regulated by stellar
feedback. In such galaxies, hydrostatic mid-plane pressure should be
balanced by the injection of kinetic energy from star formation,
either through radiation pressure or the kinetic energy from stellar
winds and supernovae remnants, or a mix of both. We compare the
star-formation intensity with the models of \citet{andrews11} and
\citet{ostriker11}, who determined which region in this diagram should
be populated by an Eddington-limited starburst. Feedback in this case
is dominated by radiation pressure in (optically thin and optically
thick) clouds with a range of dust opacities, $\kappa$. In both
models, clouds with gas-mass surface densities as observed in the Ruby
lie in the optically thick regime.

The relationship of \citet{ostriker11} is shown as a dark blue solid
line in Fig.~\ref{fig:sklaw} in the optically thick regime, and as
a dotted line at surface densities where the clouds are still optically
thin. The yellow dashed band in the same figure shows the range of
locations expected in the model of \citet{andrews11}, for their warm
starburst scenario. They account for ranges in gas-to-dust ratio and
dust temperature, which are akin to those we find in the Ruby. Both
agree very well with the location of the Ruby. We caution that
systematic effects can blur this result. For example, uncertainties in
the CO-to-H$_2$ conversion factor and gas excitation (the $r_{43/10}$
factor in Sect.~\ref{ssec:schmidtkennicutt}) make gas mass surface
density estimates uncertain by a few tenths of a~dex. The accuracy of our data
therefore does not allow us to distinguish between the detailed
assumptions of each feedback model. Nonetheless, the range covered by
radiation pressure and mechanical feedback from star formation in this
diagram is large enough that our basic result, namely, that the Ruby
falls well within the regime dominated by feedback, is robust in spite
of these uncertainties.

Likewise, the Ruby falls into the self-limited regime if stellar winds
and supernovae are the main channel through which kinetic energy is
injected into the gas. We use Starburst99 \citep{leitherer99} to
estimate the energy and momentum injection rates for a starburst with
an age of 10~Myr (Sect.~\ref{ssec:kinenergy}), and equate the pressure
created by the star formation with the hydrostatic midplane pressure.
The resulting relationship is shown as the two solid dark green lines
in the right panel of Fig.~\ref{fig:sklaw}. The lower line is the
limit for galaxies where 100\% of the kinetic energy is thermalized
into the wind, the upper line is for galaxies where this happens with
an efficiency of 50\%. Both estimates provide very similar limits to
those implied by radiation pressure, suggesting that these mechanisms
are about equally important in the most intensely star-forming
systems. We stress that radiation pressure and winds should be
considered as two co-existing, not competing processes, because the
photon flux from the young stellar population is emitted in parallel
to the mechanical energy and momentum carried by outflows from
supernovae and young stars. A somewhat steeper relationship is
produced when using the star-formation law of \citet{faucher13}, who
combined the momentum input from radiation pressure, winds, and
supernovae in an ansatz that explicitly accounts for the formation of
giant molecular clouds in a marginally Toomre-stable disk supported by
the turbulent pressure from the starburst itself. Fig.~\ref{fig:sklaw}
shows that the Ruby falls into the regime expected from a
self-regulated starburst for all these approaches.

\subsection{Star-formation efficiency}
\label{ssec:sfefficiency}

Our high-resolution data of the Ruby also allow us to estimate the
star-formation efficiency on scales that are comparable to those
probed in nearby galaxies \citep[e.g.,][]{bigiel08}. As argued by
\citet{daddi10} and \citet{genzel10}, high redshift starburst
galaxies fall above the ridge line of main sequence galaxies in the
Schmidt-Kennicutt diagram, which could imply that they convert
their gas into stars at higher efficiencies
\citep[e.g.,][]{hodge15,usero15}.  We now estimate the
star-formation rate per free-fall time to investigate if the
position of the Ruby well above the main sequence is also
matched by higher star-formation efficiencies per free-fall time.

The free-fall time, $t_{ff}$, is set by $t_{ff}=\sqrt{3 \pi / 32 G
  \rho}$, where $G$ is the gravitational constant and $\rho$ the gas
density. We estimate the three-dimensional gas density from the
observed range of (projected) gas mass surface densities of approximately
$10^{4.1-4.5}$ M$_{\odot}$ pc$^{-2}$, by assuming cloud sizes between
50~pc and 100~pc, corresponding to their Jeans lengths
(Sect.~\ref{ssec:regions12}). Resulting densities are between
$8.5\times 10^{-21}$ and $4.3\times 10^{-20}$ g cm$^{-3}$. This
suggests free-fall times between $3$ and $7\times 10^5$ yrs. Over
such timescales, approximately $150-1400$ M$_{\odot}$ pc$^{-2}$ of stars form,
corresponding to a range of approximately $1-9$\% of the available
gas-mass. This does not suggest significantly higher star-formation
efficiencies per free-fall time than in more moderately star-forming
galaxies, and is consistent with the theoretical arguments of
\citet{krumholz12}, who attributed the offset between the two
populations in the Schmidt-Kennicutt diagram to differences in the
formation and survival time of molecular clouds rather than intrinsic
offsets in the efficiency of converting gas into stars. Star-formation
in the Ruby is therefore consistent with the presence of a universal
star-formation law out to the most intense starbursts.  

\section{Gas energetics and turbulent support} 
\label{sec:limitsf}

\subsection{Kinetic energy and momentum}
\label{ssec:kinenergy}

The previous section shows that the Ruby falls into a regime of
gas-mass surface density and star-formation intensity, where star
formation should be self-regulated through the energy and momentum
injection from young stellar populations into the gas. Whether this
occurs primarily through radiation pressure or ejecta from supernovae
and young massive stars, it should leave an imprint on the gas
kinematics, through driving turbulence and perhaps causing
outflows. We will in the following analysis use our kinematic maps of
the Ruby to quantify the possible effect of feedback from star
formation onto the gas kinematics. This is particularly
interesting here, because the Ruby is the most extreme starburst
observed today with resolutions of $\la 100$ pc in the source plane.

We will now use the gas kinematics to further constrain the impact of
feedback. We note that we have not found clear evidence of outflows in
our data; this could be due to the faintness of the line emission as
seen in our high-resolution data. Spectra which do show outflow
signatures, typically also show a prominent systemic component, which
dominates the overall line profile. Moreover, it is possible that a
wind component would predominantly be in lower column density gas,
which would not necessarily lead to bright line emission
\citep[][]{sturm11}, or in lower density, and perhaps atomic rather
than molecular gas \citep[e.g.,][]{hayward15, nesvadba11}. We also
caution that the blue and redshifted components in the integrated
spectra of individual counter images, which could be taken as line
wings from outflowing gas in unresolved spectra
(Fig.~\ref{fig:familyspec}), arise from multiple gas clouds, as seen
in Fig.~\ref{fig:maps}.

\begin{figure*}
\centering
\includegraphics[width=0.7\textwidth]{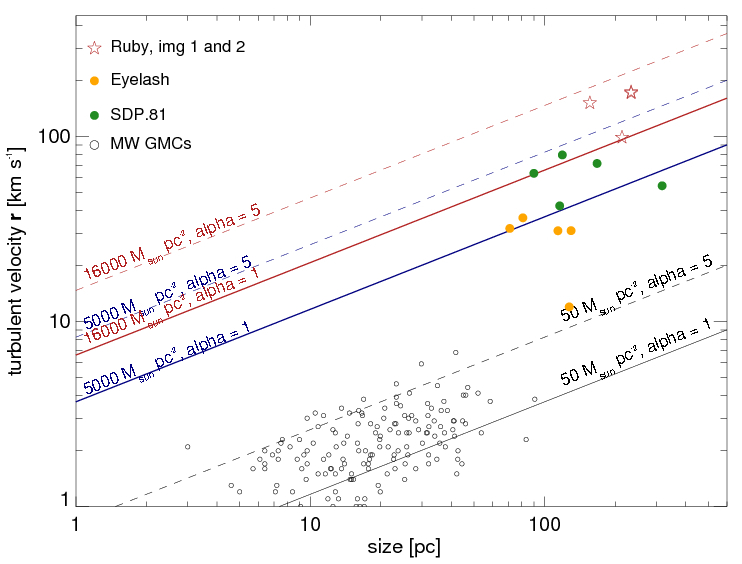}
\caption{Turbulent scaling (``Larson'') relationship
  \citep[][]{larson81} between gas velocity dispersion and cloud size
  in the source plane. Empty red stars show the sizes and line widths
  of the redshifted and blueshifted components seen in the Ruby within
  images~\#1 and \#2, respectively. Green and yellow filled circles
  show similar observations of SDP.81 and the Eyelash from
  \citet{swinbank15} and \citet{swinbank11}. Empty black circles show
  giant molecular clouds in the Milky Way. We also show expected
  line-width size relationships for gas mass surface densities,
  $\Sigma_{\rm gas} = 50$~M$_{\odot}$ pc$^{-2}$ characteristic for the
  Milky Way, and for 5000~M$_{\odot}$ and $16000$~M$_{\odot}$
  pc$^{-2}$, which correspond to the lowest and highest
  values for the Ruby, respectively (Fig.~\ref{fig:sklaw}). For each
  density we show the relationship for virial parameters $\alpha_{\rm
    vir}=1$ and 5, that is, clouds where turbulent energy either balances
  gravitational binding energy (solid line), or where it exceeds it by
  a factor~5 (dashed-dotted lines).}
\label{fig:alphavir}
\end{figure*}

We used the maps of CO surface brightness, velocity dispersion, and
gravitational magnification to estimate the kinetic energy, $E_{\rm
  kin,turb}$, and momentum, $p$, corresponding to each spatial pixel
in regions~1 and 2, by setting $E_{\rm kin,turb}=3/2\ \Sigma\ m_{\rm
  i}\ \sigma_{\rm i}^2$ and $p=\Sigma\ m_{\rm i}\ \sigma_{\rm i}$,
where $m_{\rm i}$ and $\sigma_{\rm i}$ are the molecular gas mass and
Gaussian velocity dispersion measured in each individual pixel. We
find gas kinetic energies of $E_{\rm turb,\#1.1}=(1.8\pm0.5) \times
10^{57}$ erg s$^{-1}$ and $(2.5\pm0.3) \times 10^{57}$ erg s$^{-1}$ in
regions~1 and~2, respectively, and momenta of $p_{\rm
  \#1.1}=(7.0\pm1.8)\times 10^{49}$~dyn~s and $p_{\rm
  \#2.1}=(6.3\pm0.6)\times 10^{49}$~dyn~s, respectively (all values
are corrected for gravitational magnification).

Given that the Ruby is strongly obscured in the rest-frame UV and
optical, we have no direct constraints on the stellar population,
which makes it difficult to infer the age of the starburst. Our main
purpose here, however, is to investigate whether or not the star formation is
sufficient to power the observed gas kinematics. For this work, a
lower limit on the age of the stellar population is sufficient, and
this can be obtained from the ratio of far-infrared luminosity and GHz
radio continuum. \citetalias{canameras15} showed that the Ruby falls
near the ratio $q=2.4$ typical for low-redshift galaxies, which
implies that star formation has already reached an equilibrium between
the production of massive stars and the explosion of supernovae
\citep[][]{bressan02}. This equilibrium is reached approximately $1$ to a few
times $10^7$ yrs after the onset of star formation
\citep[][]{bressan02}. Adopting a lower limit on the starburst age in
the Ruby of 10~Myrs, this suggests that we need energy injection rates
of approximately $6\times 10^{42}$ erg s$^{-1}$ and $8\times 10^{42}$ erg
s$^{-1}$ to explain the kinetic energy in images~\#1.1 and \#2.1, and
approximately $2.1\times 10^{35}$ dyn and $3.3\times 10^{35}$ dyn in momentum
injection rate.

In the following, we rely on Starburst99 \citep[][]{leitherer99}
to estimate the energy and momentum injection rates from supernovae
and stellar winds. For continuous star formation over a few times
$10^7$ yrs and solar metallicity, with a Chabrier IMF, they find
mechanical luminosities of $10^{41.8}$ erg s$^{-1}$ for each solar
mass of star formation. This corresponds to a momentum injection rate
of approximately $2.1\times 10^{33}$ dyn per solar mass formed. For image
\#1.1 with an intrinsic, magnification-corrected ${\rm SFR}=335\pm2$
M$_{\odot}$ yr$^{-1}$, this implies an energy injection rate of
$(2.1\pm0.2)\times 10^{44}$ erg s$^{-1}$ and a momentum deposition
rate of $(7.0\pm0.6)\times 10^{35}$ dyn. For image \#2.1, intrinsic
energy and momentum injection rates are $1.5\pm0.2\times 10^{44}$ erg
s$^{-1}$ and $4.9\pm0.5\times 10^{35}$ dyn, respectively. From
radiation pressure, we expect $1.5\times 10^{36}$ dyn in image~\#1.1,
and $1.1\times 10^{36}$ dyn in image~\#2.1.

We follow \citet{murray05} to estimate the momentum injection rate
from radiation pressure by setting $L_{\rm SB}/c\sim 4.6\times 10^{33}
\epsilon_3$ SFR dyn. Note that we used a low-mass cutoff of
1~M$_{\odot}$ instead of $0.1~M_{\odot}$ \citep[as would be implied by
  the Salpeter initial mass function adopted by][]{murray05}, which
approximates the Chabrier initial mass function we adopted earlier,
and increases our momentum estimate by a factor~2.3 compared to that
of \citet{murray05}.

\subsection{Turbulent support}

The above suggests that mechanical feedback from supernovae and
stellar winds can provide the kinetic energy in the gas, and either
these sources or radiation pressure can also provide sufficient
momentum to explain the observed broad line widths. As a complementary
test, we now investigate whether the observed line widths are
also consistent with those expected to keep the gas marginally
gravitationally bound, as is found in giant molecular clouds
\citep[e.g.,][]{krumholz05}.

A central quantity for turbulence-regulated star formation is the
virial parameter, $\alpha_{\rm vir}$, the ratio of turbulent and
gravitational binding energy, $\alpha_{\rm vir}=5\sigma_{t}^2 /
\pi\ G\ R\ \Sigma_{\rm gas}$, where $\sigma_{\rm t}$ is the Gaussian
line width, $G$, the gravitational constant, $R$ the cloud radius, and
$\Sigma_{\rm gas}$ the gas-mass surface density
\citep[][]{bertoldi92}. Clouds are gravitationally bound if
$\alpha_{\rm vir}<1$. Star-forming giant molecular clouds in the Milky
Way typically have $\alpha_{\rm vir}\ga 1$, that is, they are marginally
bound \citep[e.g.,][]{heyer09}.
  
In Fig.~\ref{fig:alphavir} we show where the Ruby falls relative to
the line-width size relationship of giant molecular clouds that are
kept marginally gravitationally stable by supersonic turbulence. We
show the location of clouds with $\alpha_{\rm vir}=1$ and $\alpha_{\rm
  vir}=5$, virial parameters that roughly straddle the velocity
dispersions of giant molecular clouds in the Milky Way. We show this
relationship for these virial parameters at three different gas mass
surface densities, at $50$ M$_{\odot}$ pc$^{-2}$ as found in the Milky
Way \citep[][]{heyer09}, and at 5000 and 16000 M$_{\odot}$ pc$^{-2}$,
respectively, which represent the lower and higher ranges of gas-mass
surface densities found in the Ruby. The Ruby falls into the region
expected for gas-mass surface densities as observed.

Fig.~\ref{fig:alphavir} resembles the line-width size diagrams shown by
\citet{swinbank11} for the Cosmic Eyelash and by \citet{swinbank15}
for SDP.81, and we also show these two galaxies for comparison
here. The offsets of these two galaxies from the local line-width size
relationship of nearby giant molecular clouds are somewhat smaller
than that of the Ruby, consistent with their somewhat lower gas-mass
surface densities. The more extreme position of the Ruby in the
Schmidt-Kennicutt diagram suggests that the same equilibrium
conditions hold up to the highest star-formation intensities expected
for maximal starbursts.

\citeauthor{swinbank11} also interpreted these offsets in terms of the
hydrostatic mid-plane pressure of the disks in these galaxies, arguing
that the higher pressures in high-redshift galaxies are at the origin
of the higher star-formation intensities. This is not in contradiction
with our work; however, by parameterizing this offset in terms of
$\alpha_{\rm vir}$, Fig.~\ref{fig:alphavir} highlights that it is the
balance between turbulent and hydrostatic pressure that determines the
regime where molecular clouds are marginally gravitationally
stable. Star-forming clouds are not only found for a single
equilibrium pressure, but scatter around a range of ratios of
turbulent and gravitational energy. Generally speaking, the clouds in
the Ruby do not seem to span a wider range than nearby GMCs. This
shows that the interaction between the energy and momentum injection
from star formation and the depth of the potential wells in
high-redshift galaxies together set the efficiency at which stars
form, an efficiency that appears to vary much less than the diversity
of the environments in which it occurs might suggest.

\section{Summary}
\label{sec:summary}

We presented an analysis of ALMA extended-baseline observations of
CO(4--3) and the 3-mm continuum in the $z=3.0$ dusty starburst galaxy
PLCK~G244.8$+$54.9, the ``Ruby'' at 0.1\arcsec\ spatial resolution,
probing spatial scales down to approximately 60~pc in the source plane. This
galaxy is the brightest of \textit{Planck}'s Dusty GEMS, a set of
exceptionally bright high-redshift galaxies discovered with the
\textit{Planck} and \textit{Herschel} satellites.  The Ruby forms a small
Einstein ring around a distant, massive galaxy at $z=1.52$ (C17),
magnifying two regions of the same galaxy into two sets of multiple
images. Differential lensing is not a concern for the present
analysis, however, we used the high spatial resolution to infer what
the impact of differential lensing between gas and dust would be for
unresolved data, finding shifts of approximately 10-30\% at most, much less
than what is sometimes proposed based on toy models (but the effects
in other wavebands might of course be more important).

These images sample two individual regions within a single galaxy that
are apart by roughly 470~pc in the source plane, and are spatially
resolved along, and also perpendicular to the lensing direction, with
sizes of 1.4~kpc$\times$0.7~kpc, and 1.2~kpc$\times$0.6~kpc,
respectively. Velocity gradients within individual images are larger
than the velocity difference between the two regions. Interpreting these
velocity gradients as rotational motion, we find mass surface
densities of a few $10^{10}$ M$_{\odot}$ kpc$^{-2}$, and a Toomre
parameter $Q=1.0\pm0.3$, consistent with a critically Toomre-stable
disk on large scales. 

The high star-formation intensities of up to 2200~M$_{\odot}$~yr$^{-1}$~kpc$^{-2}$ 
are clearly in the range of a maximal starburst,
and the location of the Ruby in the Schmidt-Kennicutt diagram suggests
that star formation is self-regulated by the energy and momentum
injection from radiation pressure (``Eddington-limited starburst'')
and supernova and wind ejecta from young stellar populations. Both
contribute by very similar amounts to balancing the hydrostatic
mid-plane pressure. The star-formation efficiency is approximately 1-10\%, and
we have no reason to suspect that the star-formation law as found in
the Milky Way would not be universal even in the most intense
starbursts.

We investigate whether kinetic energy from star formation could also explain 
the gas kinematics, which is a direct probe of the physical link between 
the energy output from young stars and the processes that determine the 
star-formation efficiency in scenarios of turbulence-regulated star formation. 
We find that both radiation pressure and supernovae and winds can provide 
sufficient feedback to keep molecular clouds in the Ruby marginally
gravitationally stable, akin to star-forming clouds in the Milky Way,
although at much higher mass surface densities and turbulent
pressures. Additional feedback from an AGN is not necessary, in
agreement with the absence of a bright AGN in the infrared and radio.

\section*{Acknowledgements}

We thank the anonymous referee for comments that helped improve the
paper. We also thank the ALMA Regional Center in Grenoble for their
support with preparing these observations and the ALMA staff in Chile
for carrying out the observations. RC would like to thank Claudio
Grillo for useful discussions and comments. NPHN acknowledges support
through a JAO visitor grant, and wishes to thank the ALMA staff in
Vitacura for interesting scientific discussions and their great
hospitality during an extended stay in early 2016. She also wishes to
thank Matt Lehnert for having pointed out that the Kennicutt relation
of star formation and FIR luminosity has initially been derived for a
Salpeter, not a Chabrier IMF. ML acknowledges CNRS and CNES for
support. This paper makes use of the following ALMA data:
ADS/JAO.ALMA\#2015.1.01518.S. ALMA is a partnership of ESO
(representing its member states), NSF (USA) and NINS (Japan), together
with NRC (Canada), NSC and ASIAA (Taiwan), and KASI (Republic of
Korea), in cooperation with the Republic of Chile. The Joint ALMA
Observatory is operated by ESO, AUI/NRAO and NAOJ. I.O. acknowledges
support from the European Research Council in the form of the Advanced
Investigator Programme, 321302, {\sc cosmicism}.

\bibliography{g244p8}

\end{document}